\documentclass[fleqn,usenatbib]{mnras}

\usepackage{newtxtext,newtxmath}

\usepackage[T1]{fontenc}
\usepackage{ae,aecompl}

\usepackage{graphicx}	
\usepackage{amsmath}	
\usepackage{amssymb}	

\title[Aging stars clusters with RSGs]{On aging star clusters using red supergiants independent of the fraction of interacting binary stars.}

\author[J.J. Eldridge et al.]{
J. J. Eldridge,$^{1}$\thanks{E-mail: j.eldridge@auckland.ac.nz)}
Emma R. Beasor$^{2,3}$\thanks{Hubble Fellow}
N. Britavskiy$^{4,5}$
\\
$^1$Department of Physics, University of Auckland, Private Bag 92019, Auckland, New Zealand\\
$^2$NSF National Optical-Infrared Astronomy Research Laboratory, 950 N. Cherry Ave., Tucson, AZ 85719, USA\\
$^{3}$Astrophysics Research Institute, Liverpool John Moores University, 146 Brownlow Hill, Liverpool L3 5RF, UK\\
$^{4}$Instituto de Astrof\'{i}sica de Canarias, E-38205 La Laguna, Tenerife, Spain\\
$^{5}$ Universidad de La Laguna, Dpto. Astrof\'{i}sica, E-38206 La Laguna, Tenerife, Spain\\
}

\date{Accepted XXX. Received YYY; in original form ZZZ}

\pubyear{2015}

\begin{document}
\label{firstpage}
\pagerange{\pageref{firstpage}--\pageref{lastpage}}
\maketitle

\begin{abstract}
We use the Binary Population and Spectral Synthesis (BPASS) models to test the recent suggestion that red supergiants can provide an accurate age estimate of a co-eval stellar population that is unaffected by interacting binary stars. Ages are estimated by using both the minimum luminosity red supergiant and the mean luminosity of red supergiants in a cluster. We test these methods on a number of observed star clusters and find our results in agreement with previous estimates. Importantly we find the difference between the ages derived from stellar population models with and without a realistic population of interacting binary stars is only a few 100,000 years at most. We find that the mean luminosity of red supergiants in a cluster is the best method to determine the age of a cluster because it is based o the entire red supergiant population rather than using only the least luminous red supergiant.
\end{abstract}

\begin{keywords}
binaries: general -- stars: supergiants -- stars: massive --  galaxies: star clusters: general
\end{keywords}

\section{Introduction}

Determining the age of a resolved stellar cluster is a difficult process that has many uncertainties. One important and still unclear issue is the effects of rotation \citep[e.g][]{2019A&A...622A..66G} and interacting binaries \citep[e.g.][]{1998A&A...334...21V}. Especially in changing the distribution of stars around the main-sequence turn-off, the primary feature used in age estimation. Both rotation and binaries allow stars to linger on the main sequence and be more luminous than would be expected from a simple non-rotating stellar model. The effect is obvious in old clusters as \textit{blue stragglers} are clearly separated from the main-sequence turn-off while in younger clusters the situation is less clear.

{The contribution of both rotation and binary interactions to the main-sequence turn-off will be complex, and may also be age dependent. For example recent studies have shown that there is a link between position of a star in the main-sequence turn-off and it's rotation velocity \citep[e.g.][]{2019ApJ...876..65L,2019ApJ...876..113S,2019ApJ...883..182S}. But trying to separate the importance of stellar rotation relative to binary interactions is difficult given that the most rapidly rotating stars most likely arise from binary interactions as described by \citet{DeMink2013}. Therefore while there is clearly a link between rotation and the distribution of stars in the main-sequence turn-off, it is not clear if the distribution of rotation velocities is linked to the initial rotation velocity distribution or related to the population of interacting binaries within the stellar population. It would be useful if there was another method to determine the age of a star cluster that was not strongly affected by either binary interactions nor stellar rotation.}

Recently both \citet{Beasor} and \citet{Britavskiy} have suggested a novel and accurate way to estimate the ages of star clusters by using the least luminous, and effectively the oldest, red supergiant (RSG). For these stars the age of the star is greater than the time it spends in the evolutionary phase which provides a tight age constraint. They also suggested that binary interactions such as mass transfer and mergers would only give rise to more luminous \textit{red straggler} RSGs, relatively younger stars, that can be ignored.

Due to the significant implications of such a method providing a new insight into star cluster ages, as well as allowing future investigations into impact of interacting binaries on the turn-off star population, this idea should be tested further. In this letter we investigate this idea using state-of-the-art theoretical stellar population models, from the BPASS (Binary Population and Spectral Synthesis) project, v2.2.1 \citep{2017PASA...34...58E,2018MNRAS.479...75S}. This code allows us to fully investigate the effect binary stars have on using RSGs on age determination.

\section{Stellar population models}

\begin{figure*}
	\includegraphics[width=1.95\columnwidth]{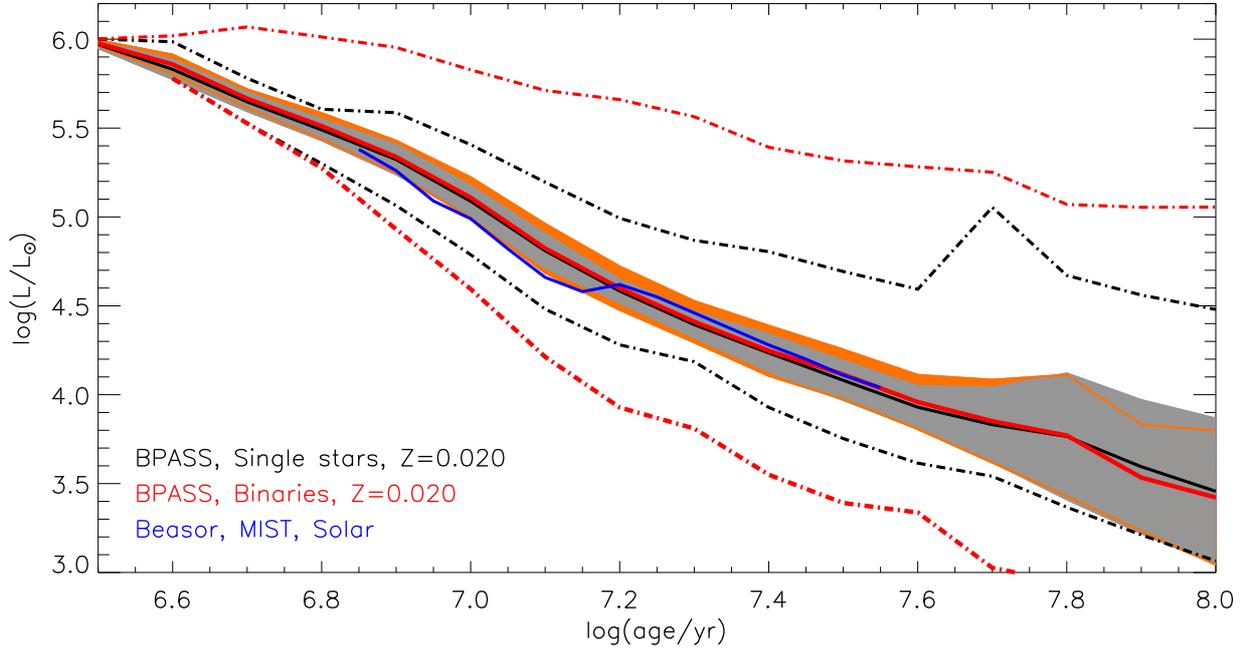}
	\caption{Bolometric RSG luminosity versus time at $Z=0.020$. Solid red lines binary RSG mean luminosity and 1$\sigma$ range is shown by the orange shaded region. Black solid lines are the same for single star populations, with grey shading. The dash-dotted black lines are the maximum and minimum RSG luminosities for the single star population while the red dash-dotted lines are the same values for the binary population. The blue solid lines are the mean minimum RSG luminosity calculated by \citet{Beasor}. }
    \label{fig:bolometric}
\end{figure*}

We use the Binary Population and Spectral Synthesis (BPASS) v2.2.1 models \citep[see][for full details]{2017PASA...34...58E,2018MNRAS.479...75S} to synthesize RSG populations. We determine the minimum, maximum, mean luminosities and the standard deviation about the mean luminosity of RSGs versus age. Importantly we do this for a population made solely of single stars and another that incorporates a realistic interacting binary star population based on the results of \citet{2017ApJS..230...15M}. 

We use results from three metallicities, with the metal mass fractions, $Z=0.004$, 0.008 and 0.020. These are suitable for the Small Magellanic Cloud (SMC), Large Magellanic Cloud (LMC) and our Galaxy, respectively. We use the fiducial BPASS initial mass function with a minimum mass of 0.1$M_{\odot}$, a slope of $dN/dM \propto -1.30$ up to 0.5$M_{\odot}$ with a slope of $dN/dM \propto -2.35$ up the maximum mass of 300M$_{\odot}$.

To determine the nature of the RSG population at each BPASS logarithmic time bin of width 0.1 dex we record the luminosities of red supergiants, defined as stars that have completed core hydrogen burning and have a surface temperature of $\log(T_{\rm eff}/K)\le3.66$. We search for the minimum luminosity, the maximum luminosity and calculate the mean luminosity and the standard deviation.

We note that in star clusters it is unlikely that we observe RSGs at the minimum possible luminosity we predict in BPASS. \citet{Beasor} pointed out that RSGs evolve rapidly through those evolutionary phases. \citet{Beasor} therefore calculated a most likely minimum luminosity RSG by using the RSG luminosity distribution and sampling the lowest luminosity RSG, assuming 50 RSGs in a cluster. We have tested this method but found the result calculated is consistent with the luminosity given at the mean RSG luminosity minus the standard deviation of the population. This is reasonable as only 16 per cent of RSGs would be below this limit. Furthermore the most probable minimum luminosity of any RSGs will be close to this lower standard deviation. In this work, for simplicity, we take the most likely minimum RSG luminosity to be the luminosity defined by the distribution mean minus the distribution's standard deviation at each time.

We show our derived distribution luminosity parameters for our RSG populations in Figure \ref{fig:bolometric}. We can see that the minimum luminosities are close for single star and binary populations at early times but diverge at older ages. The binary population maximum luminosity is always significantly above that from single stars due to red stragglers. In comparison the mean luminosities calculated are very close between single and binary populations. These luminosities also appear to provide a good age estimate up to approximately 100 Myrs. Although beyond approximately 40~Myrs the distribution of expected luminosities increases due to increasing number of asymptotic-giant branch (AGB) stars in the stellar populations.

As noted above, in Figure \ref{fig:bolometric} the minimum and maximum luminosities versus age are either lower or higher respectively for the binary population relative to the singe-star population. The higher maximum is due to mergers and mass transfer making more luminous red supergiants at older ages than expected from a single star population as suggested by \citet{Britavskiy} and \citet{Beasor}. The lower minimum luminosities are at odds with the expected assumption that the minimum luminosity RSGs are single stars. In fact RSGs in a binary that fill their Roche lobe and lose mass all decrease in luminosity, this is a consistent prediction from binary evolution models \citep[e.g][]{2001A&A...369..939W}. Thus including such interacting stars causes the minimum possible red supergiant luminosity to be lower and thus whether a red supergiant is interacting with a companion should be checked. This will be observationally challenging. However these lower luminosity binary stars are rare and the method employed by \citet{Beasor} our using a most likely minimum luminosity still works. Although the standard deviation for the binary population is greater than that of the single-star population.

In Figure \ref{fig:bolometric} we have overplotted the mean minimum luminosity calculated by \citet{Beasor}. We can see that these lie on the lower 1-$\sigma$ line for our population up to 16~Myrs after which the luminosity matches our mean luminosity line. The difference between the \citet{Beasor} and the BPASS lines provide an estimate of the systematic uncertainty that occurs due to our choice of stellar models. \citet{Beasor} used the MIST stellar models \citep{2016ApJS..222....8D} and we have compared our single star models of the same initial mass to those to determine the reason for the difference. The initial mass range where the change occurs in the \citet{Beasor} relation is between 13 to 14$M_{\odot}$.

A detailed comparison between the BPASS and MIST models is beyond the scope of this work. However, we have examined the models to understand where there might be a difference of the order of 0.1~dex in the expected minimum luminosities. Above 13~M$_{\odot}$ the stellar tracks are quite similar, although the BPASS models reach lower luminosities by 0.1~dex at the cool side of the Hertzsprung gap than the MIST models. But the BPASS models also reach higher luminosities by 0.1~dex at the end of their evolution than the MIST tracks so on average the mean luminosities are the same. The RSG lifetimes also appear similar which suggests that the models, even with minor differences, agree to first order.

At 13~M$_{\odot}$ and below blue loops (the movement on the Hertzsprung-Russell diagram during helium burning for lower mass stars) occur in some of the stellar tracks but again during the entire Hertzsprung gap model from MIST are 0.1~dex less luminous than the BPASS tracks. Then during the RSG evolution the lifetime of the MIST models is less than that of the BPASS models by a few hundred thousand years. The MIST models also tend to reach the RSG before the BPASS models which together equates to older ages for a RSG of the same luminosity. The differences lead to the bump in the minimum luminosity line at 16~Myrs.

We expect the stellar model have this behaviour due to the MIST tracks changing certain details around the MESA stellar models at this mass range. In comparison, BPASS models make no change to the numerical or physical details with stellar mass and this is represented by significantly smoother age--luminosity relations. As we have stated above this does give an idea of the systematic uncertainty implicit in using an assumed stellar evolution model. We note that the BPASS tracks both single and binary have already been extensively validated against many observations of stars, e.g. \citet{2017PASA...34...58E} and \citet{2019PASA...36...41E}.

In light of all the above outlining the sensitivity of the most likely minimum luminosity RSG to many factors we suggest another method to estimate a stellar population age with RSGs. This is to use the mean RSG luminosity calculated from all the RSG in a cluster. This removes some (but not all) of the uncertainties between stellar models and importantly uses all the RSGs in a cluster rather than relying on the details of a single star. To evaluate how useful this method and that of the using minimum luminosity RSG are, and how dependent on binary fraction the results are, we analyse the clusters discussed in \citet{Beasor} and \citet{Britavskiy} and other similar clusters for which data on the RSG population exists.

\section{Ages of observed clusters}

We show the ages derived from RSGs for Galactic, LMC and SMC clusters with RSGs in Table \ref{tab:rsg}. We note that one of these clusters, Upper Sco, contains only one RSG, Antares. We see the age estimates using both the most-likely minimum luminosity and mean luminosity lie within the range of previous estimates. The ages from the most-likely minimum luminosity RSG tend to be older than those estimated from the mean RSG luminosity.  

The differences between the age estimates from our single star and binary star populations using the mean RSG luminosity is typically a few 100,000 years, much smaller than the ages derived and within the uncertainty. This again reinforces the consistency of the idea that RSGs can be used to give useful age constraints on the age of stellar populations. The different ages from the single star and binary star populations for the most-likely minimum luminosity estimate are typically a few times greater than that from the mean RSG luminosity. This confirms the hypothesis that a binary population independent age estimate is possible by using the mean RSG luminosity in a cluster. Using the most likely minimum luminosity RSG may produce a small overestimate of the age. However the values still agree within the calculated uncertainties.

{We note that for the most-likely minimum luminosity ages we have derived we have not given the uncertainty in this value. They should be at least the same magnitude as those from the mean luminosity but we expect these to be higher. One reason is that the slope of the trends of luminosity in age in Figure \ref{fig:bolometric} are similar and thus a similar error in the luminosity will give the same error in the age. Another reason is that depending on just one star to estimate the age we must consider whether we are truly sampling the RSG luminosity distribution well enough to have a star at the most-likely minimum luminosity. To estimate the uncertainty we would need to look at the RSG luminsity distribution at each age, sampling this with the number of observed RSGs to see what the distribution of luminosities we derive for the lowest luminosity RSG would be, as in \citet{Beasor}. However this requires a separate model for each individual cluster introducing extra computational cost.}

{Most of the clusters in our sample have ages in the range of 10 to 20~Myrs. We note for clusters beyond this the discrepancy between the most-likely minimum luminosity age and mean luminosity age are greatest. With the former overestimating the age significantly for several of the clusters. This again demonstrates the problematic nature of relying on a single star to derive an age.}.

The errors we present here are significant when considering the age on a linear scale as they are of the order of 0.15 dex or 40 per cent. The uncertainty in the luminosity of the RSGs is generally small, less than 0.1 dex. The same is true for the error in the mean luminosity (taken to be $\sigma/\sqrt{N}$). While we see that for the standard deviation of the mean luminosities derived from the BPASS models are also similar around 0.1 dex. Combining these uncertainties implies that there is always a minimum accuracy for any age estimate.

\begin{figure*}
	\includegraphics[width=1.8\columnwidth]{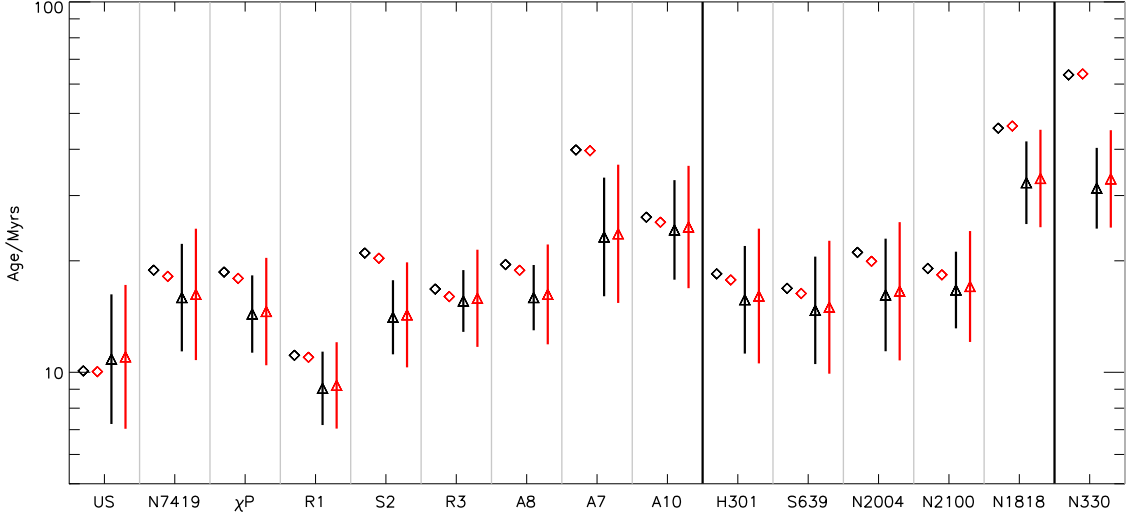}
	\caption{Comparison of the ages of the star clusters listed in Table \ref{tab:rsg} using either the mean minimum luminosity RSG (diamonds) or the mean RSG luminosity (triangles), derived using single star BPASS models (black) or the binary star BPASS models (red).} 
    \label{fig:compare}
\end{figure*}

\begin{table*}
	\centering
	\caption{The ages of a number of Galactic (first 5), LMC (second 5), SMC (last 1) clusters estimated by using the minimum and mean luminosity of RSGs. The data is taken from \citet{Beasor}, \citet{Britavskiy}, Beasor et al. (in prep.). For the previous age estimate we list the range and the best value from previous work that includes \citet{keller1,keller2,2009A&A...498..109C,2010A&A...513A..74N,2010ApJS..186..191C,2013A&A...552A..92M,2019A&A...629A.117O}. Clusters indicated with an asterisk are ages calculated on the RSGs in the clusters that have been spectroscopically confirmed.}
	\label{tab:rsg}
	\begin{tabular}{lccccccccc} 
		\hline
		\hline
                &  &  &   &   &  \multicolumn{2}{c}{$L_{\rm min}$} &  \multicolumn{2}{c}{$L_{\rm mean}$ }  \\
		 &Number & & & Previous &  \multicolumn{2}{c}{estimate / Myr} &  \multicolumn{2}{c}{estimate / Myr}\\
		Cluster &of RSGs &$\log(L_{\rm min}/L_{\odot})$ & $\log(L_{\rm mean}/L_{\odot})$& Estimate / Myr & Single & Binary &  Single & Binary \\
		
		\hline
		Upper Sco & 1 &  4.99   & 4.99$\pm$0.15    & 11   &              10.1      & 10.0     &  10.8$_{-3.6}^{+5.4}$   &    11.0$_{-4.0}^{+6.2}$\\
		NGC 7419  & 5 &  4.37  &  4.58$\pm$0.13     &  7.1--21    &      18.9     &  18.2   &    15.9$_{-4.5}^{+6.3}$  &     16.2$_{-5.4}^{+8.2}$\\
		$\chi$-Per & 8 &  4.38  & 4.68$\pm$0.07       &  7.9-22   &      18.6     &  17.9   &    14.4$_{-3.1}^{+3.9}$  &   14.6$_{-4.1}^{+5.8}$\\
		RSGC1 & 14 & 4.87  &  5.19$\pm$0.05   &              14    &     11.1      & 11.0     &  9.0$_{-1.8}^{+2.3}$    &   9.2$_{-2.2}^{+2.8}$\\
		Stevenson 2(RSGC2) & 26 & 4.28  & 4.70$\pm$0.06  &  14-20    &     21.0      & 20.3    &   14.1$_{-2.9}^{+3.6}$   &    14.3$_{-4.0}^{+5.5}$\\
		RSGC3*             & 9  &  4.47  &   4.60$\pm$0.05 & 16-20   &       16.8 &      16.0     &  15.6$_{-2.7}^{+3.3}$    &   15.8$_{-4.1}^{+5.6}$\\
		Alicante 8 (RSGC4)*  & 8  &  4.34  &   4.58$\pm$0.06 &    16-20 &    19.5 &      18.9    &   15.9$_{-2.9}^{+3.6}$    &   16.2$_{-4.3}^{+5.9}$\\
		Alicante 7*         & 7  &3.81   &  4.29$\pm$0.13  & --  &       39.9 &       39.7  &     23.2$_{-7.2}^{+10.3}$  &     23.7$_{-8.3}^{+12.7}$\\
		Alicante 10*        &  4 &  4.1  &  4.26$\pm$0.09  &   --    &   26.2  &     25.4  &     24.2$_{-6.5}^{+8.8}$    &   24.7$_{-7.8}^{+11.4}$\\
		\hline  
		Hodge 301 & 4  & 4.46   &  4.70$\pm$0.11  & 3--24      &      18.4     &  17.8   &    15.7$_{-4.5}^{+6.2}$   &    16.1$_{-5.5}^{+8.4}$\\
		SL 639 & 4   &     4.54   & 4.77$\pm$0.11   & 7--22     &     16.9  &     16.3    &   14.7$_{-4.2}^{+5.8}$     &  15.0$_{-5.1}^{+7.7}$\\
		NGC 2004 & 6  &    4.35  & 4.67$\pm$0.12  & 6.3--24     &      21.1  &     19.9   &    16.2$_{-4.8}^{+6.8}$     &  16.5$_{-5.8}^{+8.9}$\\
		NGC 2100 & 18  &    4.43 &  4.64$\pm$0.05 & 7.1--22     &      19.1   &    18.3   &    16.7$_{-3.5}^{+4.5}$     &  17.0$_{-5.0}^{+7.0}$\\
		NGC 1818 & 13  & 3.803  &   4.14$\pm$0.06 &   25         &   45.6     &  46.2     &  32.5$_{-7.4}^{+9.5}$     &  33.4$_{-8.7}^{+11.8}$\\
		\hline
		NGC 330  & 14  & 3.642  &  4.20$\pm$0.07  &   32        &     63.5    &   63.9  &     31.4$_{-7.0}^{+9.0}$    &   33.3$_{-8.7}^{+11.8}$\\
		\hline  
	\end{tabular}
\end{table*}

\section{Discussion \& Conclusions}

\begin{figure*}
	\includegraphics[width=1.9\columnwidth]{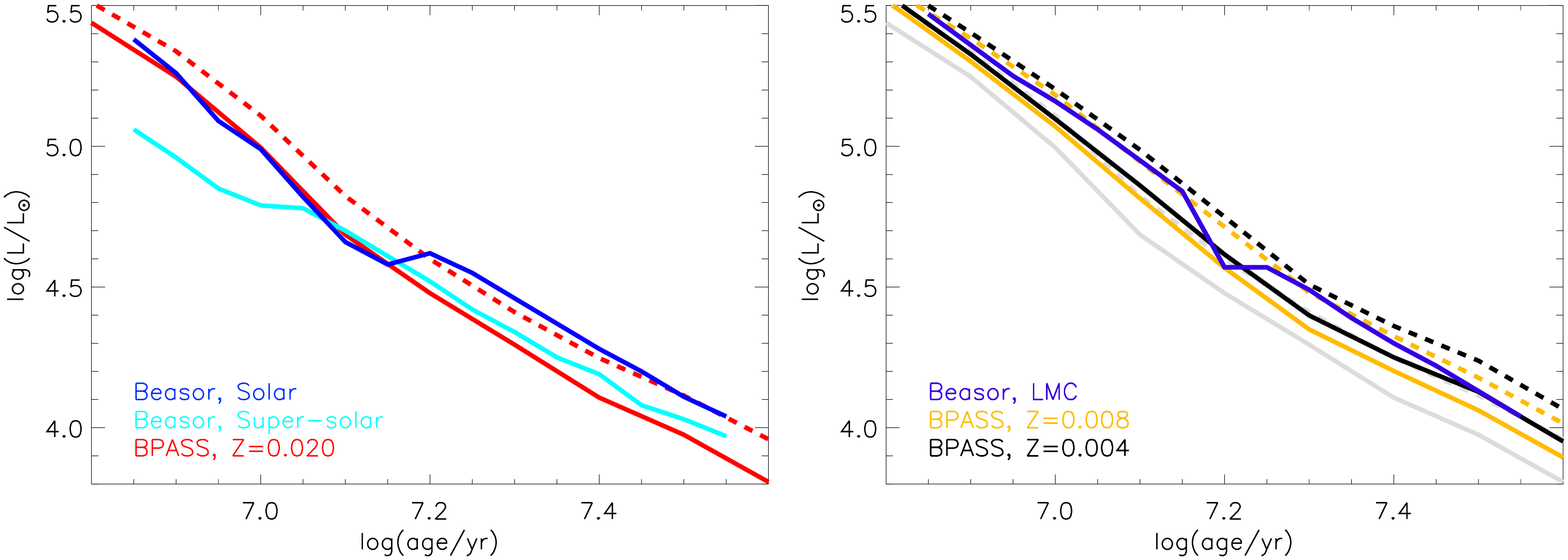}
	\caption{Comparison of the BPASS mean minimum RSG luminosities with those calculated by \citet{Beasor}. The solid lines are the mean minimum RSG luminosities and the dashed lines are mean RSG luminosities. The left-hand panel is for Solar metallicity and above and the right-hand panel is for sub solar metallicities. The grey lines in the right-hand panel are the BPASS Z=0.020 lines. All BPASS models are the stellar populations including binary stars.}
    \label{fig:bolometric2}
\end{figure*}

We have used the BPASS stellar population models to investigate how interacting binary stars affect the accuracy of estimating star cluster ages from either using the most-likely minimum luminosity RSG and the mean RSG luminosity. While stellar mergers and mass gainers in stellar binaries will become more massive and more luminous RSGs than expected for a stellar population's age, it was suggest by \citet{Britavskiy} and \citet{Beasor} that the minimum luminosity red supergiants are most likely to be the result of evolution of an effectively single star. We found that less luminous RSGs can be formed by interacting binaries than expected from single star evolution. However, such stars will be rare and thus the most probable minimum luminosity RSG method of \citet{Beasor} is still applicable.

In Figure \ref{fig:bolometric2} we compare all our different relations at different metallicities as well as those calculated by \citet{Beasor}. The relations vary due to metallicity and from using different stellar models. This suggests we must be wary that there is a systematic uncertainty in any age estimate from using the RSGs that is dependent on the assumptions in the stellar models, whether that be the mass-loss rates, mixing scheme applied or the initial metallicity.

However, while there is an offset, all the age-luminosity relations in Figure \ref{fig:bolometric2} have a similar gradient. We note the relations from BPASS tend to be smoother than those calculated by the MIST models. These shapes of the relations will be dependent on the assumptions assumed in the MIST stellar models \citet{2016ApJS..222....8D}, most likely at stellar masses below 20M$_{\odot}$. The BPASS relations are smoother as these models use consistent physical and numerical ingredients used over the full mass range of the BPASS models.

{One effect that we have not considered in detail is stellar rotation. \citet{Beasor} did consider how this affected the ages derived from the minimum luminosity RSG. They found that the ages between rotating models could be older than non-rotating models by as much as 10 per cent. Stellar rotation extending the main sequence lifetime by mixing in fresh hydrogen into the core during the main sequence. Furthermore, with binary populations the situation is made complex due to the fact that the most rapidly rotating stars arise from binary interactions \citep{DeMink2013}. Importantly, not every RSGs we observe in a population is likely to a star that had rotated rapidly enough to impact on its evolution. Therefore by using the mean luminosity of all RSGs in a population possible biasing of the age, due to the lowest luminosity RSG being the product of a rapidly rotating star, is reduced. Given the small difference in ages found by \citet{Beasor} compared to the typical uncertainties we find from the distribution of RSG luminosities in our population we suggest that rotation should not be an important factor in deriving ages from the mean RSG luminosity.}

Given our findings we conclude that the best method to use RSGs to estimate the age of a star cluster is to use all the observed RSGs in a cluster to estimate a mean RSG luminosity. BPASS results indicate that the difference  between the mean luminosity for a single star only or a realistic binary star population are to within 0.05 dex over a significant age range. Therefore the estimated age is only weakly dependent on the inherent binary population of a star cluster as postulated by \citet{Beasor} and \citet{Britavskiy}.

We have included in on-line supplementary information a Table \ref{tab:example_table} that contains our derived mean RSG luminosities from our BPASS models so others can easily use this method to derive the ages of star clusters. The future uses of such a method are many. First, by constraining the age of the star cluster from the RSGs the turn off stars can be examined in detail. Specifically it can be estimated how many stars are beyond the turn-off expected for a single star population and thus how the interacting binary population and stellar rotation alter the turn-off appearance. Secondly, the stellar populations around some supernova progenitors can also have their age estimated accurately. By using all coeval RSGs alone, rather than the apparent main-sequence turn-off, a significantly tighter age and thus initial mass constrains could be achieved.

\section*{Acknowledgements}
The authors thank the referee for useful feedback that improved the clarity of the paper. The authors thank H\'{e}lo\"{i}se Stevance and Max Briel for proof reading of the manuscript.
JJE acknowledges travel funding and support from the University of Auckland. Support for this work was provided by NASA through Hubble Fellowship grant HST-HF2-51428 awarded by the Space Telescope Science Institute, which is operated by the Association of Universities for Research in Astronomy, Inc., for NASA, under contract NAS5-26555. NB acknowledges support
from the Spanish Government Ministerio de Ciencia, Innovaci\'on y
Universidades through grants PGC-2018-091\,3741-B-C22 and from the
Canarian Agency for Research, Innovation and Information Society
(ACIISI), of the Canary Islands Government, and the European Regional
Development Fund (ERDF), under grant with reference ProID2017010115.

\label{lastpage}
\end{document}



\begin{table*}
	\centering
	\caption{BPASS estimates for the minimum, mean and maximum luminosities for stellar populations with metallicity of $Z=0.020$, 0.008 and 0.004.}
	\label{tab:example_table}
	\begin{tabular}{ccccccc} 
	\hline
$Z=0.020$	& \multicolumn{3}{l}{Single stars} & \multicolumn{3}{l}{Binary stars} \\
$\log({\rm age/yrs})$ &  $\log(L_{\rm min}/L_{\odot})$ &  $\log(L_{\rm mean}/L_{\odot})$ 	& $\log(L_{\rm max}/L_{\odot})$ &  $\log(L_{\rm min}/L_{\odot})$ &  $\log(L_{\rm mean}/L_{\odot})$ &	 $\log(L_{\rm max}/L_{\odot})$ \\ 
	\hline
      6.5   &    5.95  &     5.97  $\pm$    0.02   &     6.00 &      5.96  &     5.98 $\pm$     0.02 &      6.00\\
      6.6   &    5.78  &     5.83 $\pm$     0.06    &   5.99  &    5.78 &      5.86 $\pm$     0.05   &    6.02\\
      6.7   &    5.53  &     5.65 $\pm$    0.05     &  5.78   &  5.53    &   5.67   $\pm$   0.05     &  6.07\\
      6.8   &    5.30  &     5.49 $\pm$     0.06    &   5.61  &    5.27   &    5.51 $\pm$     0.07   &    6.01\\
      6.9   &    5.06  &     5.32 $\pm$     0.08    &   5.59  &   4.93    &   5.34  $\pm$    0.09    &   5.95\\
      7.0   &    4.79  &     5.09 $\pm$     0.09    &   5.41  &   4.59    &   5.11  $\pm$     0.11   &    5.83\\
      7.1   &    4.48  &     4.81 $\pm$     0.09    &   5.20  &   4.21    &   4.82   $\pm$    0.14   &    5.71\\
      7.2   &    4.28  &     4.58 $\pm$     0.06    &   4.99  &    3.93   &    4.60  $\pm$     0.12  &     5.66\\
      7.3   &    4.18  &     4.39  $\pm$    0.07    &   4.87  &   3.81    &   4.41  $\pm$     0.12   &    5.56\\
      7.4   &    3.93  &     4.24  $\pm$     0.11   &    4.80 &    3.55   &    4.25 $\pm$     0.14   &    5.39\\
      7.5   &    3.75  &     4.08  $\pm$     0.11   &    4.69 &    3.39   &    4.11 $\pm$      0.14  &     5.32\\
      7.6   &    3.62  &     3.93  $\pm$     0.12   &    4.59 &    3.34   &    3.96 $\pm$      0.15  &     5.28\\
      7.7   &    3.54  &     3.83 $\pm$      0.21   &    5.05 &   3.03    &   3.85   $\pm$    0.23   &    5.25\\
      7.8   &    3.37  &     3.77  $\pm$     0.35   &    4.67 &   2.94    &   3.77   $\pm$    0.34  &     5.07\\
      7.9   &    3.21  &     3.60  $\pm$     0.37   &    4.56 &      2.90 &      3.53  $\pm$     0.30  &     5.05\\
      8.0   &    3.06  &     3.46$\pm$      0.41    &   4.48  & 2.78    &    3.42 $\pm$      0.38   &    5.06\\
   \hline
 $Z=0.008$	& \multicolumn{3}{l}{Single stars} & \multicolumn{3}{l}{Binary stars} \\
$\log({\rm age/yrs})$ &  $\log(L_{\rm min}/L_{\odot})$ &  $\log(L_{\rm mean}/L_{\odot})$ &	 $\log(L_{\rm max}/L_{\odot})$ &  $\log(L_{\rm min}/L_{\odot})$ &  $\log(L_{\rm mean}/L_{\odot})$ &	 $\log(L_{\rm max}/L_{\odot})$ \\ 
  \hline
    6.4   &     --      &   --                   &--      &    6.75   &    6.81   $\pm$      0.11     &  6.88\\
      6.5   &    6.03     &  6.05 $\pm$       0.02  &     6.09  &  6.42  &     6.45 $\pm$        0.33   &    6.84\\
      6.6   &    5.85     &  5.89 $\pm$       0.06  &     6.06  &  5.85   &    5.93 $\pm$        0.10   &    6.61\\
      6.7   &    5.63     &  5.71 $\pm$       0.05  &     5.83  &   5.61   &    5.70 $\pm$       0.06   &    6.10\\
      6.8   &    5.45     &  5.57 $\pm$       0.07  &     5.80  &   5.38    &   5.57  $\pm$      0.05   &    6.10\\
      6.9   &    5.26     &  5.38 $\pm$       0.08  &     5.63  &   5.17    &   5.38  $\pm$      0.082  &     5.98\\
      7.0   &    5.00     &  5.17 $\pm$       0.09  &     5.45  &   4.86    &   5.18  $\pm$       0.11  &     5.87\\
      7.1   &    4.77     &  4.93  $\pm$      0.09  &     5.26  &    4.53   &    4.94 $\pm$        0.13  &     5.71\\
      7.2   &    4.36     &  4.69  $\pm$      0.10  &     5.09  &  4.07     &  4.71 $\pm$        0.15    &   5.70\\
      7.3   &    4.26     &  4.47  $\pm$      0.08  &     4.90  & 3.97      & 4.48   $\pm$      0.13     &  5.56\\
      7.4   &    4.02     &  4.31  $\pm$      0.09  &     4.84  &  3.74     &  4.33 $\pm$        0.12    &   5.37\\
      7.5   &    3.85     &  4.16 $\pm$       0.09  &     4.75  &  3.64     &  4.18   $\pm$      0.12    &   5.27\\
      7.6   &    3.70     &  3.98  $\pm$      0.10  &     4.67   &  3.46    &   4.02  $\pm$       0.12   &    5.09\\
      7.7   &    3.57     &  3.86  $\pm$       0.11 &      5.10  &    3.36  &     3.90 $\pm$        0.14 &      5.16\\
      7.8   &    3.47     &  3.78  $\pm$       0.24 &      5.08  &   3.25   &    3.74 $\pm$        0.19   &    5.08\\
      7.9   &    3.34     &  3.71  $\pm$       0.37 &      4.65  &    3.07  &     3.68 $\pm$        0.36  &     5.10\\
      8.0   &    3.21     &  3.64  $\pm$       0.43 &     4.57   &  2.86    &   3.56   $\pm$      0.37   &    5.06\\

 
		\hline
		 $Z=0.004$	& \multicolumn{3}{l}{Single stars} & \multicolumn{3}{l}{Binary stars} \\
$\log({\rm age/yrs})$ &  $\log(L_{\rm min}/L_{\odot})$ &  $\log(L_{\rm mean}/L_{\odot})$ &	 $\log(L_{\rm max}/L_{\odot})$ &  $\log(L_{\rm min}/L_{\odot})$ &  $\log(L_{\rm mean}/L_{\odot})$ &	 $\log(L_{\rm max}/L_{\odot})$ \\ 
  \hline
    6.3   &    6.88   &    6.90 $\pm$       0.02   &    6.91  &   6.96    &   7.00  $\pm$      0.04   &    7.08\\
      6.4   &    6.74   &    6.85 $\pm$       0.05   &    6.93 &    6.71    &   6.89  $\pm$      0.09   &    7.16\\
      6.5   &    6.07   &    6.10 $\pm$       0.03   &    6.15 &   6.71     &  6.75  $\pm$       0.26   &    7.04\\
      6.6   &    5.91   &    5.93  $\pm$      0.05   &    6.09 &    5.91    &   6.05 $\pm$        0.28  &     6.89\\
      6.7   &    5.69   &    5.74 $\pm$       0.05   &    5.87 &    5.66    &   5.74 $\pm$       0.06   &    6.15\\
      6.8   &    5.52   &    5.58 $\pm$       0.06   &    5.78 &     5.51   &    5.60 $\pm$       0.06  &     6.13\\
      6.9   &    5.33   &    5.41 $\pm$       0.08   &    5.67 &    5.22    &   5.40  $\pm$      0.08   &    6.00\\
      7.0   &    5.09   &    5.19 $\pm$       0.08   &    5.45 &    4.98    &   5.20 $\pm$        0.11  &     5.88\\
      7.1   &  4.89     &  4.97 $\pm$      0.09     &  5.26  &  4.71    &   4.99  $\pm$       0.13  &     5.79\\
      7.2   &    4.61   &    4.73  $\pm$      0.10   &    5.13  &   4.41  &     4.75  $\pm$       0.13  &     5.75\\
      7.3   &    4.34   &    4.49 $\pm$       0.07   &    4.92  &    4.07  &     4.51 $\pm$        0.11 &      5.54\\
      7.4   &    4.08   &    4.35  $\pm$      0.08   &    4.86  &  3.87 &      4.36  $\pm$       0.11    &   5.40\\
      7.5   &    3.91   &    4.20  $\pm$      0.08   &    4.78  &   3.73  &     4.24 $\pm$        0.11   &    5.28\\
      7.6   &    3.77   &    4.02  $\pm$      0.09   &    4.70  &   3.59   &    4.07  $\pm$       0.11   &    5.18\\
      7.7   &    3.65   &    3.88  $\pm$      0.09   &    4.58  &   3.42    &   3.93   $\pm$      0.12   &    5.06\\
      7.8   &    3.56   &    3.78  $\pm$       0.13  &     5.12 &    3.27   &    3.78  $\pm$       0.12  &     5.12\\
      7.9   &    3.46   &    3.67   $\pm$      0.26  &     5.09 &    3.15   &    3.70   $\pm$      0.28  &     5.09\\
      8.0   &    3.35   &    3.66   $\pm$      0.42  &     4.63 &   3.05    &   3.60  $\pm$       0.33   &    5.08\\
   
		\hline
	\end{tabular}
\end{table*}